\documentclass[prb,twocolumn,superscriptaddress,floatfix,showpacs,amsmath,amssymb]{revtex4-2}

\usepackage{dcolumn}
\usepackage{bm}
\usepackage{graphicx,color}
\usepackage{epstopdf}
\usepackage{xspace}
\usepackage[english]{babel}
\usepackage{verbatim}
\usepackage{braket}
\usepackage{booktabs}
\usepackage{ulem} 
\usepackage[version=3]{mhchem}
\usepackage{amsmath,mathtools}

\newcommand{\rk}[1]{\textcolor{black}{#1}}

\renewcommand{\arraystretch}{1.3}

\newcommand{\TN}{$T_\mathrm{N}$}

\newcommand{\NdB}{NdB$_4$}
\newcommand{\DLL}{$dL/L$}

\newcommand{\tn}{$T_\mathrm{N}$}
\newcommand{\tit}{$T_\mathrm{IT}$}
\newcommand{\tlt}{$T_\mathrm{LT}$}
\newcommand{\mb}{$\mu_{\rm B}$}

\newcommand{\jmk}{J/(mol\,K)}

\begin{document}
	
\title{Magnetic phase diagram and magneto-elastic coupling of NdB$_4$ studied by high-resolution capacitance dilatometry up to 35~T}

	\author{R. Ohlendorf}~\thanks{These authors contributed equally to this work}
	\affiliation{Kirchhoff Institute for Physics, Heidelberg University, 69120 Heidelberg, Germany}
	
	\author{S. Spachmann}~\thanks{These authors contributed equally to this work}
	\affiliation{Kirchhoff Institute for Physics, Heidelberg University, 69120 Heidelberg, Germany}
		
	\author{L.~Fischer}
	\affiliation{Kirchhoff Institute for Physics, Heidelberg University, 69120 Heidelberg, Germany}

    \author{F.~L.~Carstens}
	\affiliation{Kirchhoff Institute for Physics, Heidelberg University, 69120 Heidelberg, Germany}
	
	\author{D.~Brunt}
	\affiliation{Department of Physics, University of Warwick, Coventry, CV4 7AL, United Kingdom}
	\affiliation{National Physical Laboratory, Teddington, TW11 0LW, United Kingdom}
	
    \author{G.~Balakrishnan}
    \affiliation{Department of Physics, University of Warwick, Coventry, CV4 7AL, United Kingdom}

    \author{O.~A.~Petrenko}
    \affiliation{Department of Physics, University of Warwick, Coventry, CV4 7AL, United Kingdom}

    \author{R.~Klingeler}\email{klingeler@kip.uni-heidelberg.de}
    \affiliation{Kirchhoff Institute for Physics, Heidelberg University, 69120 Heidelberg, Germany}
    	
\begin{abstract}
We report high-resolution dilatometry studies on single crystals of the Shastry-Sutherland-lattice magnet NdB$_4$ supported by specific heat and magnetometry data. Our dilatometric studies evidence pronounced anomalies at the phase boundaries which imply strong magneto-elastic coupling. The evolution of the three zero-field phase transitions separating distinct antiferromagnetic phases at \tn~$=17.2$~K, \tit~$=6.8$~K and \tlt~$=4.8$~K can thus be traced in  applied magnetic fields which provides the magnetic phase diagrams for $B\parallel c$ up to 15~T and for $B\parallel [110]$ up to 35~T. New in-field phases are discovered for both field directions and already known phases are confirmed. In particular, phase boundaries between different phases are unambiguously shown by sign changes of observed anomalies and corresponding changes in uniaxial pressure effects. For $B||c$, we find a 1/4-magnetization plateau in addition to a previously reported plateau at 1/5 of the saturation magnetization. \tn\ increases for $B\parallel c$ in fields up to 15~T implying that magnetic moments of the all-in/all-out structure in the high temperature AFM ordered phase are driven towards the $c$ axis in high magnetic fields. Uniaxial pressure dependencies ${\partial}T_{\mathrm{crit}}/{\partial}p_{\mathrm{c}}$ of the phase transition temperatures for magnetic fields and pressure applied along the $c$ axis are derived from the data.

\end{abstract}

\date{\today}
 \maketitle

\section{Introduction}

Geometric frustration can lead to the suppression or even complete inhibition of long-range magnetic order, which in turn is associated with a macroscopic degeneracy of ground states. If long-range order evolves, the competing interactions often yield complex phase diagrams with energetically similar spin configurations~\cite{Ramirez1994}. Prominent examples of geometrically frustrated lattices are the triangular, kagome and pyrochlore lattices~\cite{Shimizu2003,Shen2016,Pratt2011,Lee2005,Yan2011,Han2012,Gingras2014,Chalker1992,Gardner2010}. The rare earth tetraboride (\textit{R}B$_4$; \textit{R} = rare earth ion) family has its rare earth ions arranged on the geometrically frustrated Shastry-Sutherland lattice (SSL)~\cite{Shastry1981, Gabani2020}. As a consequence, many \textit{R}B$_4$ compounds show a number of competing phases at low temperatures~\cite{Gabani2020}. A seemingly generic feature in the \textit{R}B$_4$ compounds, which raised considerable interest, is the appearance of fractional magnetization plateaus as seen in NdB$_4$~\cite{Brunt2018}, TbB$_4$~\cite{Yoshii2008,Yoshii2009}, HoB$_4$~\cite{Matas2010}, ErB$_4$~\cite{Ye2017, Matas2010}, and TmB$_4$~\cite{Siemensmeyer2008, Trinh2018, Lancon2020}. While many frustrated systems realize quantum magnets, rare earth ions exhibit large magnetic moments which makes rare earth compounds prime systems for investigating frustrated behavior in the classical limit. Like all other \textit{R}B$_4$ compounds with trivalent rare earth ions, NdB$_4$, which is reported here, is metallic~\cite{Johnson1963,Brunt2019}.

NdB$_4$ crystallizes in a tetragonal structure belonging to the D$^5_{4h}$-P4/$mbm$ space group~\cite{Etourneau1985}. This structure can be viewed as comprising of a boron and a neodymium sublattice. The boron atoms form chains of octahedra along the $c$ direction, which are connected by two additional boron atoms and form rings in the $ab$ plane. The neodymium ions are situated above and below the centres of the boron rings, forming the SSL in the $ab$ plane. The magnetic behavior of NdB$_4$ is due to the Nd$^{3+}$ ions with the electron configuration 4f$^3$ realized in the $\prescript{4}{}{I}_{9/2}$ ground state. Previous reports have shown three successive phase transitions in zero magnetic field at 17.2, 6.8, and 4.8~K, from the paramagnetic high-temperature phase into a commensurate antiferromagnetic (cAFM) phase, an incommensurate AFM phase (referred to as intermediate temperature (IT) phase here) and a low-temperature (LT) AFM phase, respectively~\cite{Brunt2018,Watanuki2009,Ohlendorf2021}. The cAFM phase exhibits an all-in magnetic structure in the $ab$ plane with the magnetic moments slightly tilted towards the $c$ direction; the tilts  alternate in opposite direction on neighbouring sites~\cite{Metoki2018}. The electronic ground state consists of two Kramers doublets forming a pseudo quartet~\cite{Yamauchi2017}. It has been suggested that the orbital degrees of freedom contribute to the phase transition at 4.8~K, which might lead to the formation of an orbitally ordered ground state~\cite{Ohlendorf2021}.

The effect of long-range magnetic order on the NdB$_4$ lattice is studied by capacitance dilatometry at temperatures between 1.3 and 300~K and in magnetic fields up to 35~T along the [110]-direction of the crystal and up to 15~T along the $c$ direction. Pronounced thermal expansion and magnetostriction anomalies signal the phase boundaries and allow the construction of the magnetic phase diagrams. The unprecedented resolution of capacitance dilatometry enables us to confirm phase boundaries in fields up to 10~T from previous publications~\cite{Watanuki2009, Brunt2018} as well as to uncover a total of seven as-yet unreported phases both in low and high magnetic fields. Magnetization and specific heat measurements support the construction of the phase diagrams and enable us to quantify uniaxial pressure effects.

\section{Experimental Methods}

Single crystals of NdB$_4$ were grown by the optical floating-zone technique as reported in detail in Ref.~\onlinecite{Brunt2019}. The relative length changes $dL_i/L_i$ along the crystallographic [001] ($c$ axis) and [110] directions (space group 127), respectively, were studied on an oriented cuboid-shaped single crystal of dimensions $1.476 \times 1.478 \times 1.880~$mm$^{3}$. Measurements were performed in static magnetic fields up to 15~T along the [001] direction and up to 35~T along the [110] direction for varying temperature \rk{upon heating and after cooling the sample in zero magnetic field (ZFC)} (thermal expansion, $L(T, B = \rm{const})$) as well as for fixed temperatures and varying magnetic fields (magnetostriction, $L(B, T = \rm{const})$). \rk{The reference length against which the changes in length were recorded was updated at the beginning of each measurement. Thus the relative length changes are to be understood with respect to the initial length at the temperature and magnetic field values at which the respective measurement started (the zero point is always at the start of the measurement).} Magnetic fields were always applied along the measurement direction. The dilatometric measurements were carried out using three-terminal high-resolution capacitance dilatometers~\cite{Kuechler2012, Kuechler2017} and the linear thermal expansion and magnetostriction coefficients, ${\alpha_i=1/L_i\times dL_i(T)/dT}$ and ${\lambda_i=1/L_i\times dL_i(B)/dB}$, were derived. Measurements in fields up to 15~T were done in a home-built setup~\cite{Werner2017} in Heidelberg (KIP) while higher field studies were done in the high-field magnet laboratory (HFML) in Nijmegen. Static magnetization $M$ and magnetic susceptibility $\chi=M/B$ were studied up to 7~T in a Magnetic Properties Measurement System (Quantum Design MPMS3 SQUID magnetometer) and up to 14~T in a Physical Properties Measurement System (Quantum Design PPMS-14) using the vibrating sample magnetometer (VSM) option. Specific heat measurements were performed in PPMS using a relaxation method.

\section{Results}

Pronounced magneto-elastic coupling in \NdB~yields clear anomalies in the thermal expansion and specific heat at the three successive phase transitions appearing at $T_{\rm N}$=17.2(1)~K, $T_{\rm IT}$=6.8(1)~K and $T_{\rm LT}$=4.8(1)~K in zero magnetic field~\cite{Ohlendorf2021}. The anomalies show a discontinuous character at $T_{\rm LT}$ while both transitions at higher temperatures are continuous. Due to their distinctive nature the phase boundaries can be well traced to higher magnetic fields, where a number of new phases has been found for both $B \parallel c$ and $B\perp c$.

\subsection{$B\parallel c$}

\begin{figure}[htb]
	\centering
	\includegraphics [width=\columnwidth,clip] {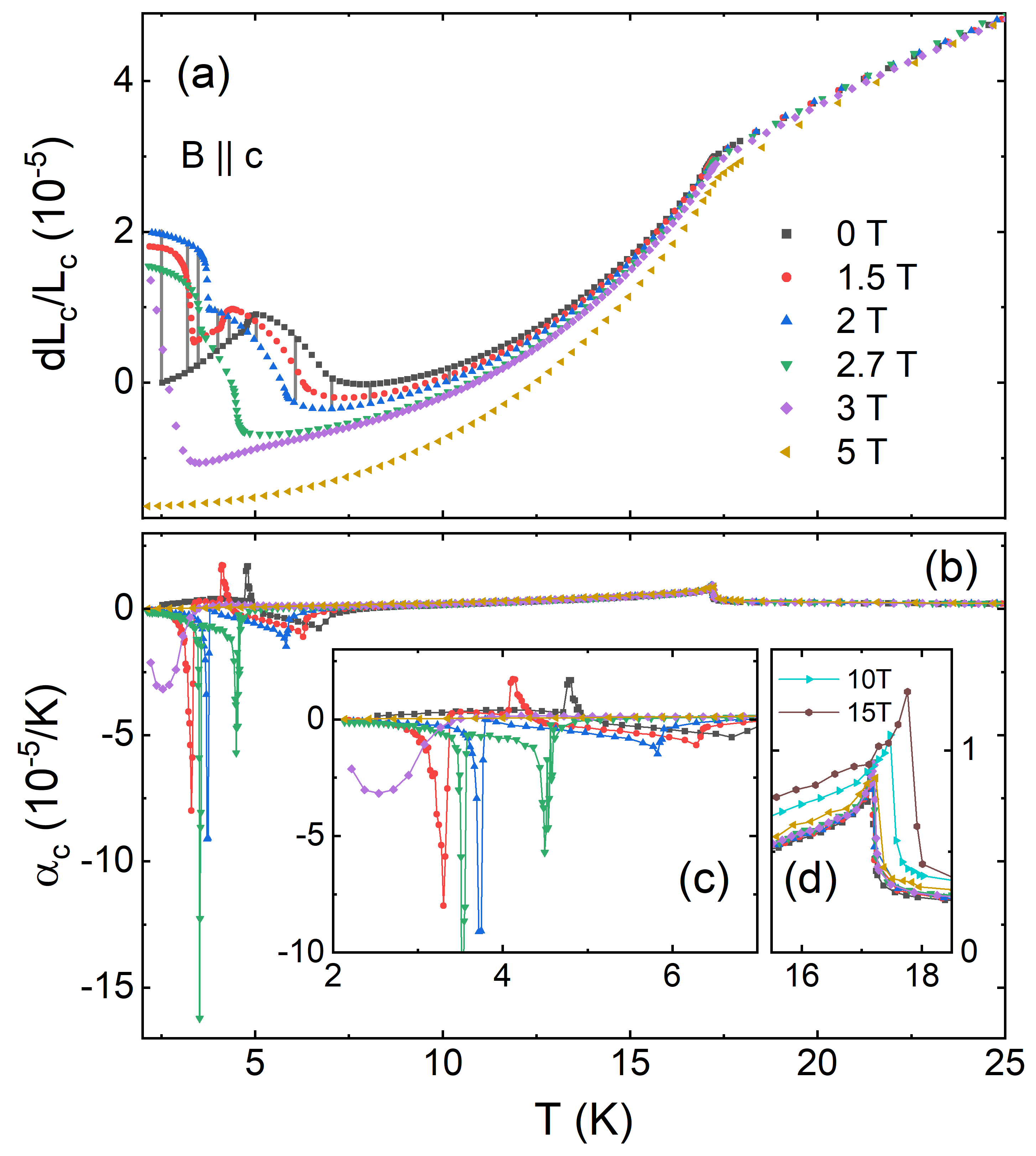}
	\caption{(a) Relative length changes and (b, c) corresponding thermal expansion coefficients for magnetic fields $B\parallel c$ in selected fields up to 5~T. (d) Thermal expansion coefficients for magnetic fields $B\parallel c$ in selected fields up to 15~T and temperatures between 15.5 and 18.5~K. The relative length changes at different magnetic fields in (a) are shifted by the measured magnetostriction at 14~K from 0~T to the respective field. Grey vertical lines for the $B=2$~T data at different temperatures exemplify the measured magnetostriction.} \label{TE_c}
\end{figure}

\renewcommand{\arraystretch}{1.2}
\begin{table*}[htb]
\setlength{\tabcolsep}{2pt}
\caption{Summary of anomalies in the thermal expansion with $B\parallel c$, including the nature of the phase boundary (discontinuous/continuous (d/c)), the sign of uniaxial pressure dependencies, and the description of general effects when increasing $B$. In the first and second column the anomalies are numbered in order of appearance and labeled according to the phase boundaries in Fig.~\ref{PD_Both}.}
\label{TE_c_pt}
\begin{center}
\begin{tabular}{c c | c c c c c }
		
\hline\hline
\# &boundary& \textit{T}(K) & \textit{B}(T) & Type& ${\partial}T/{\partial}p_c$ & behavior as $B \nearrow$\\
\hline
1& PM-HT & $17.2(1)~(T_{\rm N})$ & 0   & c& $> 0$ & shifts towards higher \textit{T} for $B \parallel c\gtrsim 6$~T \\
2   &HT-IT& $6.8(1)~(T_{\rm IT})$& 0 & c & $< 0$ & suppressed in fields\\
2a &HT-III&4.5(1) & 2.7         & d    & $< 0$ & suppressed in fields\\		
2b & HT-II &2.5(3)& 3       & d & $< 0$      & broadened peak\\
3&IT-LT & $4.8(1)~(T_{\rm LT})$& 0& d& $> 0$ & suppressed in fields\\
4&LT-I& 3.3(1)& 1.5& d& $< 0$ & transition into in-field phase\\
4a&IT-I&3.7(1)&2&d& $< 0$& shifts to higher \textit{T}\\
4b&III-II&3.5(1)&2.7& d& $< 0$& shifts to lower \textit{T}\\
\hline\hline
\end{tabular}	
\end{center}
\end{table*}
\renewcommand{\arraystretch}{1}

Thermal expansion data for selected magnetic fields $B\parallel c$ up to 15~T are shown in Fig.~\ref{TE_c}. In zero field, the above-mentioned transitions are clearly visible. Application of a magnetic field $B \parallel c$ suppresses the transition temperatures \tit\ and \tlt\ and the anomalies are no longer visible at 5~T and above. In contrast, \tn\ remains largely field-independent up to 5~T, but then noticeably increases in fields of 10 and 15 T (see Fig.~\ref{TE_c}(d)). Positive $\lambda$-like anomalies in the thermal expansion coefficient imply a positive uniaxial pressure dependence, i.e., ${\partial}T_{\rm N}/{\partial}p_c~>0$. Details are summarized in Table~\ref{TE_c_pt} (row 1). The continuous transition at \tit\ features a negative uniaxial pressure dependence, and \tit\ shifts to lower temperatures for $B>0$~T. This also holds for \tlt\ where, however, ${\partial}T_{\rm LT}/{\partial}p_c~>0$ is found. We also note that, in the paramagnetic phase at $T>T_{\rm N}$, our data indicate finite negative magnetostriction of the $c$ axis for fields applied along the \textit{c} axis at least up to 50~K, i.e., $\lambda (50~{\rm K}, 15~{\rm T})\simeq -1.2\times 10^{-6}/{\rm T}$ (see Fig.~S12 in the SM~\cite{SM}). We attribute this behaviour to short-range correlations in this temperature regime~\cite{Ohlendorf2021}.

The nature of the transitions as well as the sign of associated anomalies change in applied fields $B \parallel c$. For example, at 2.7~T a jump in the relative length changes around 4.5~K (green circles in Fig.~\ref{TE_c}(a)) indicates that the transition now shows first order character in contrast to a continuous behavior up to 2~T. This change in behavior corresponds to a transition into a newly discovered low temperature phase (phase III in our phase diagram, see Fig.~\ref{PD_Both}(a) below). Slightly increasing the field to 3~T leads to a suppression of the transition towards lower temperatures and is accompanied by a strong broadening of the peak in the thermal expansion coefficient (purple data in Fig.~\ref{TE_c}(c)). As further measurements presented below show, this anomaly belongs to a transition into another low-temperature in-field-phase (phase II). We also note an additional discontinuous transition appearing for $B \parallel c=1.5$~T at 3.3(1)~K (\#4 in Table~\ref{TE_c_pt} and phase boundary I-IT in Fig.~\ref{PD_Both}(a)). A complete summary of the observed anomalies and their main properties is presented in Table~\ref{TE_c_pt}.

\begin{figure}[htb]
	\centering
	\includegraphics [width=\columnwidth,clip] {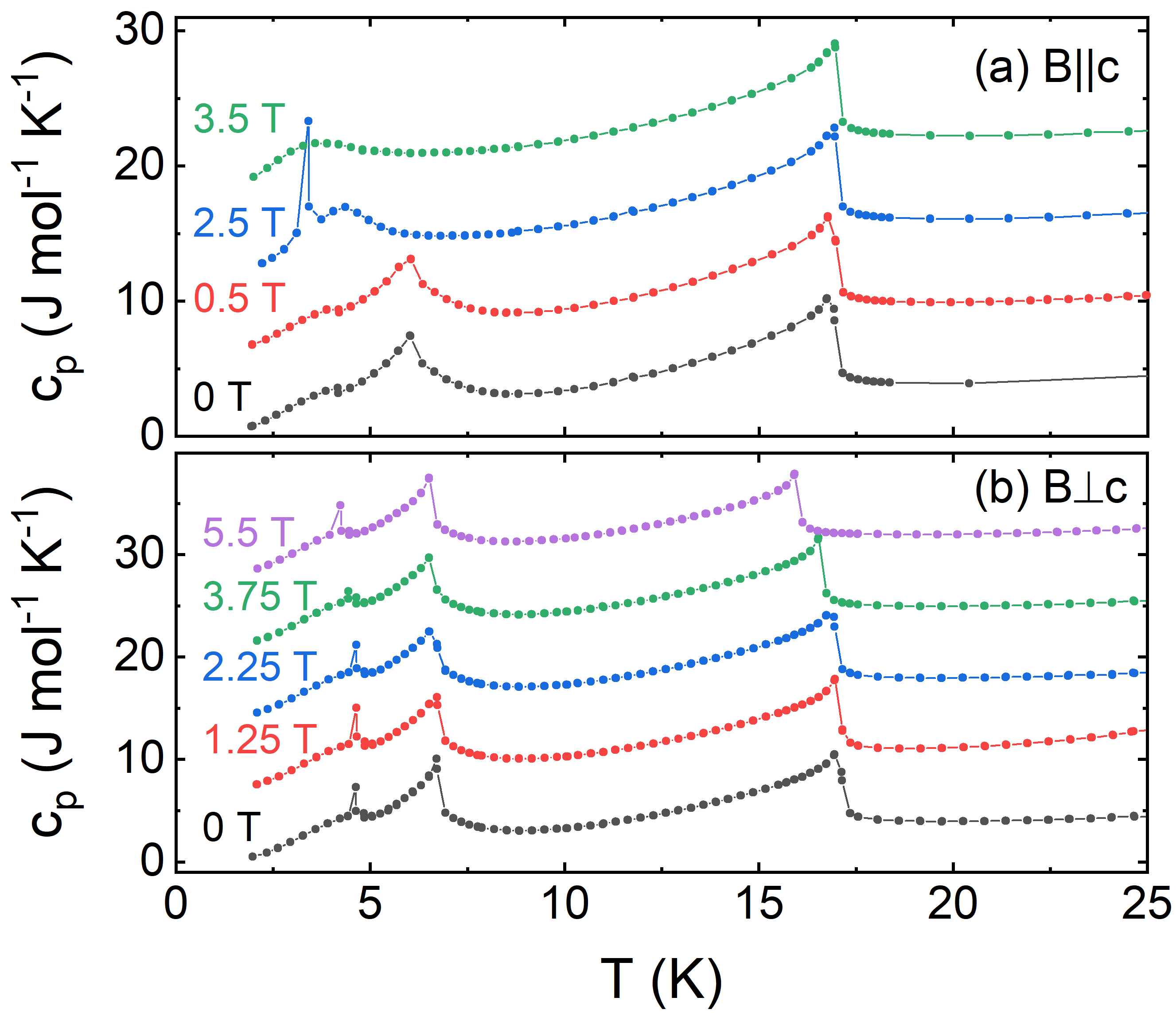}
	\caption{Specific heat for (a) $B\parallel c$ and (b) $B\perp c$. Data are offset vertically by 6/7~\jmk\ in (a/b) respectively.} \label{cp}
\end{figure}

Specific heat measurements up to 3.5~T, with $B\parallel c$, confirm the observed phase boundaries (Fig.~\ref{cp}(a)). Note that first order phase transitions are hard to resolve with the relaxation method and may thus not be visible in specific heat measurements, whereas the thermal expansion measurements are well suited to detect discontinuities and possess higher resolution.


\begin{figure}[htb]
	\centering
	\includegraphics [width=\columnwidth,clip] {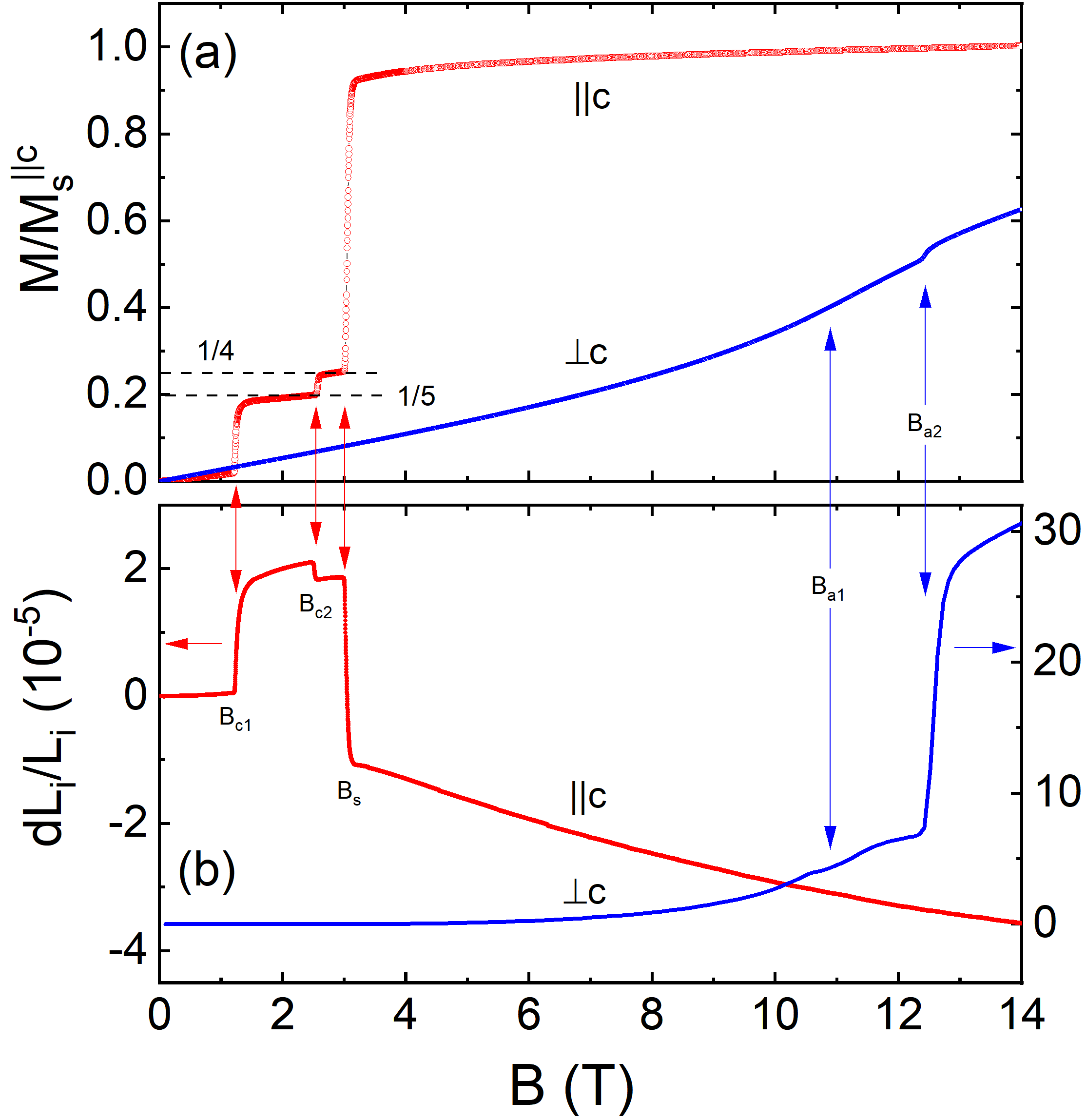}
	\caption{(a) Magnetization $M(B)$ and (b) magnetostriction $dL_{\rm i}(B)/L_{\rm i}$, at $T=2$~K, for $B\parallel c$ and $B\perp c$, respectively. All data show up-sweeps of $B$. Due to the inaccessibility of the saturation magnetization $M_{\rm s}(B\perp c)$, the magnetization data have been normalized to $M(B||c =14~\rm{T})$ for both field directions. Dashed lines signal the 1/5- and 1/4-magnetization plateaus; vertical arrows show corresponding anomalies in $M$ and $dL/L$.
 } \label{plateau}
\end{figure}

\begin{figure}[htb]
	\centering
	\includegraphics [width=0.95\columnwidth,clip] {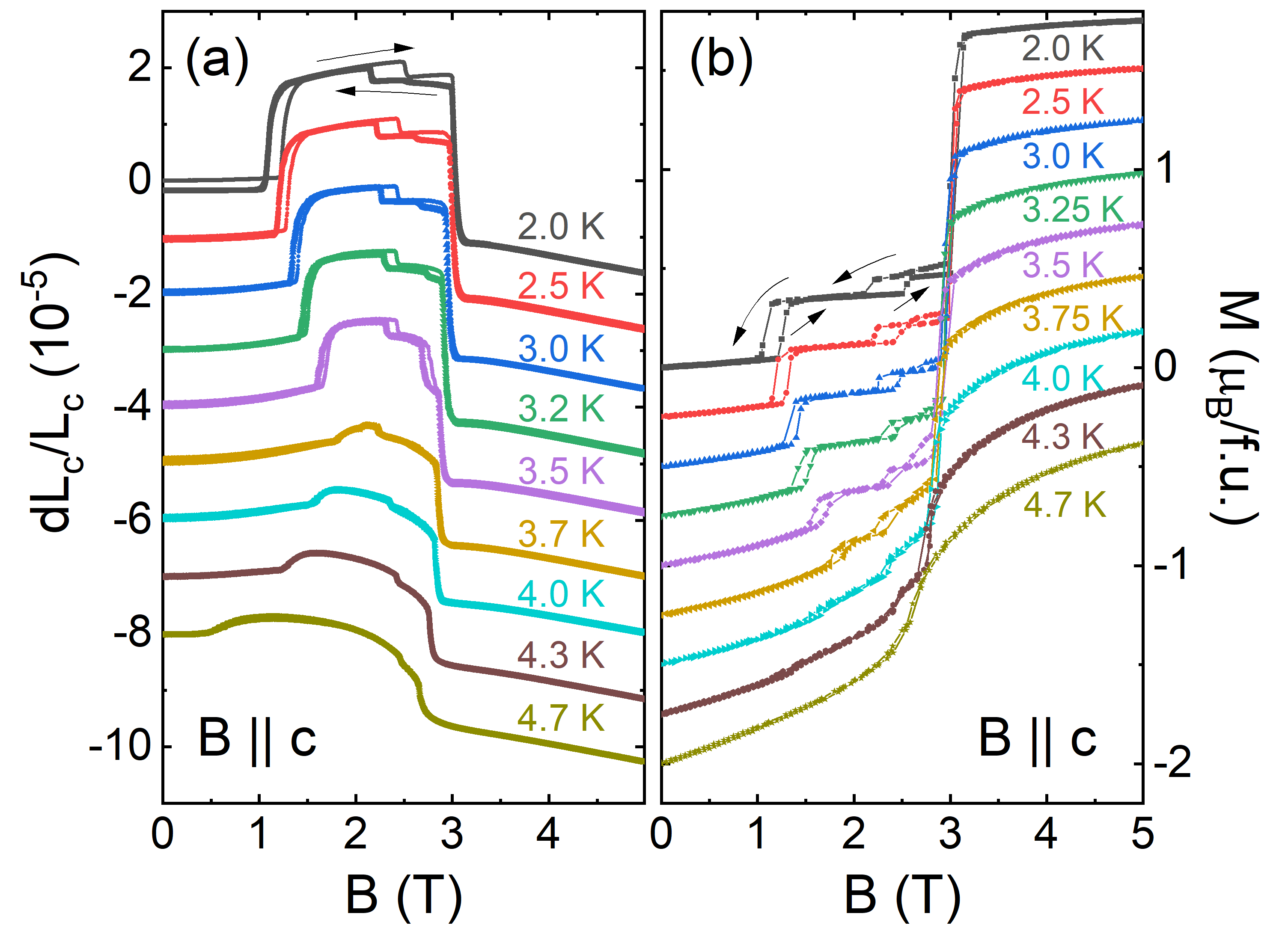}
	\caption{Comparison of (a) magnetostriction $dL(B)/L$ and (b) isothermal magnetization $M(B)$ at temperatures from 2~K to 4.7~K in fields $B \parallel c$ up to 5~T. Field sweep directions are indicated by arrows. Magnetostriction data are offset vertically by \rk{-$1\times10^{-5}$}, with up- and down-sweeps superimposed at 5~T. Magnetization data are offset by -0.25~$\mu_{\mathrm{B}}/$f.u.} \label{MS_c}
\end{figure}

In order to further illustrate magneto-elastic coupling in NdB$_4$, Fig.~\ref{plateau} shows the field dependence of the magnetization and the magnetostriction upon increasing $B$ up to 14~T, at $T=2$~K. The data display several phase transitions driven by the magnetic field which are associated with anomalies in both $M$ and \DLL . Specifically, for $B||c$ there are three successive anomalies in the magnetostriction at $B_{\rm c1}=1.1$~T, $B_{\rm c2}=2.3$~T, and $B_{\rm s}=3$~T, respectively, all of which are associated with jump-like increases in magnetization. The discontinuous character of the first two transitions is confirmed by hysteretic behavior (see Fig.~\ref{MS_c}).

The anomaly at $B_{\rm c1}$ indicates the previously observed appearance of a magnetization plateau at 1/5 of the saturation magnetization and may be attributed to a 3-up-2-down phase~\cite{Brunt2018}. At $B_{\rm c2}$, we find an additional 1/4 magnetization plateau which was not observed in the previously reported data. $B_{\rm s}$ eventually signals the onset of the magnetically polarised phase~\cite{Brunt2018}. We however note that we find $M(2~{\rm K},14~{\rm T})=1.70(5)$~\mb /f.u. which is only about 1/2 (52(2)~\%) of the full saturation moment for free Nd$^{3+}$ ions. One hence may speculate about the presence of an 1/2 magnetization plateau in the accessible field region $B_{\rm s}\leq B\leq 14$~T. Indeed, an extended 1/2 plateau is predicted by Monte-Carlo simulations on a Shastry-Sutherland lattice with multifold long-range interactions~\cite{Huo2013}. If we would consider the theoretical full moment and interpret the saturation regime above $B_{\rm s}$ as an 1/2 plateau, the plateau values at $B_{\rm c1}$ and $B_{\rm c2}$ would then be about 1/8 and 1/10, respectively, instead of 1/4 and 1/5. However, in contrast to a 1/5 plateau which is directly attributed to the 3-up-2-down configuration, there is no such straightforward explanation of a hypothetical 1/10 plateau. In addition, in pulsed high-magnetic field studies there is no significant further increase of $M(B||c)$ up to 50~T~\cite{Brunt2018} so that a hypothetical 1/2 plateau would have to extend to fields above 50~T. In contrast to $M(B||c)$, $M(B\perp c)$ has been found to steadily increase even at 50~T and a rough extrapolation suggests that the theoretical fully saturated magnetization is reached only at $B\geq80$~T.

To discuss the response of the lattice, at $B_{\rm c1}$, the  discontinuity in $M$ is associated with a jump-like increase in length along the $c$ axis when the 1/5 plateau phase evolves while the $c$ axis discontinuously shrinks both at the appearance of the 1/4 plateau phase at $B_{\rm c2}$ and at $B_{\rm s}$. The observed signs of the magnetostriction anomalies imply that the 1/5 plateau phase will be suppressed by application of uniaxial strain along the crystallographic $c$ direction as $\partial B_{\rm s}/\partial p_{\rm c}$ and $\partial B_{\rm c2}/\partial p_{\rm c}$ are negative while $\partial B_{\rm c1}/\partial p_{\rm c}>0$. \rk{Quantitatively, using~\cite{Barron-White1999}}
\begin{equation}
    \frac{\partial B_{\rm cr}}{\partial p_{\rm c}}=V_{\rm m}\frac{\Delta L_{\rm c}/L_{\rm c}}{\Delta M_{\rm c}},\label{CC}
\end{equation}
\rk{to determine the initial uniaxial pressure dependencies of the various critical fields, the experimentally observed jumps in magnetization and magnetostriction in Fig.~\ref{plateau} yield the uniaxial pressure dependencies listed in Table~\ref{tabplateau}. Here, $V_m = 3.22\times 10^{-5}$~m$^3$ is the molar volume of \NdB.~\cite{Brunt2019}}

\renewcommand{\arraystretch}{1.2}
\begin{table}[!ht]
\setlength{\tabcolsep}{4pt}
\caption{Calculated uniaxial pressure dependencies, at $T=2$~K, of critical fields $B_{\rm c1}$, $B_{\rm c2}$, and $B_{\rm s}$ obtained by means of Eq.~\ref{CC} using the data in Fig.~\ref{plateau}.\label{tabplateau}}
\centering
    \begin{tabular}{c | c | c}
        ${\partial}B_{\mathrm{c1}}/{\partial}p_{\rm c}$ & ${\partial}B_{\mathrm{c2}}/{\partial}p_{\rm c}$ & ${\partial}B_{\mathrm{s}}/{\partial}p_{\rm c}$  \\ \hline
        0.34(2)~T/GPa & -0.06(2)~T/GPa & -0.58(2)~T/GPa \\ 
    \end{tabular}
\end{table}
\renewcommand{\arraystretch}{1}

The evolution of the critical fields appearing for $B||c$, upon heating up to 4.7~K, is shown in Fig.~\ref{MS_c} (see also Fig.~\ref{PD_Both}(a) below). The main findings are as follows: (1) $B_{\rm c1}$ shifts to higher fields as the temperature is increased to 3.7~K and the hysteresis becomes less pronounced. Upon further heating, the jump vanishes and the anomaly appears more kink-like; no hysteresis is observed anymore. Additionally, the transition gets suppressed to lower fields with higher temperature; in this regime, it corresponds to the phase boundary between the low temperature (LT) phase and the intermediate temperature (IT) phase, which is of second order. (2) $B_{\rm c2}$ is rather independent on $T\leq 3.5$~K with diminishing but visible hysteresis. For $T\gtrsim 4$~K, hysteresis vanishes and there is a clear positive slope of $B_{\rm c1}(T)$. (3) No clear hysteresis is resolved at $B_{\rm s}$; the associated anomalies decrease in magnitude and get suppressed to lower magnetic fields for higher temperatures. For temperatures between 2 and 3.5~K we note a small additional jump in the magnetostriction for down-sweeps in succession of the first large jump.

\subsection{$B\parallel [110]$}

Measurements of the specific heat for $B\perp c$ up to 5.5~T are shown in Fig.~\ref{cp}(b). Three successive anomalies in zero field correspond to the phase transitions at $T_{\rm N}$, $T_{\rm IT}$ and $T_{\rm LT}$. Up to 5.5~T all anomalies get slightly suppressed towards lower temperatures, corresponding well to our thermal expansion data for low magnetic fields along the [110] axis (see Figs.~S1 to S3 in the Supplemental Material (SM)~\cite{SM}).

\begin{figure}[htb]
	\centering
	\includegraphics [width=\columnwidth,clip] {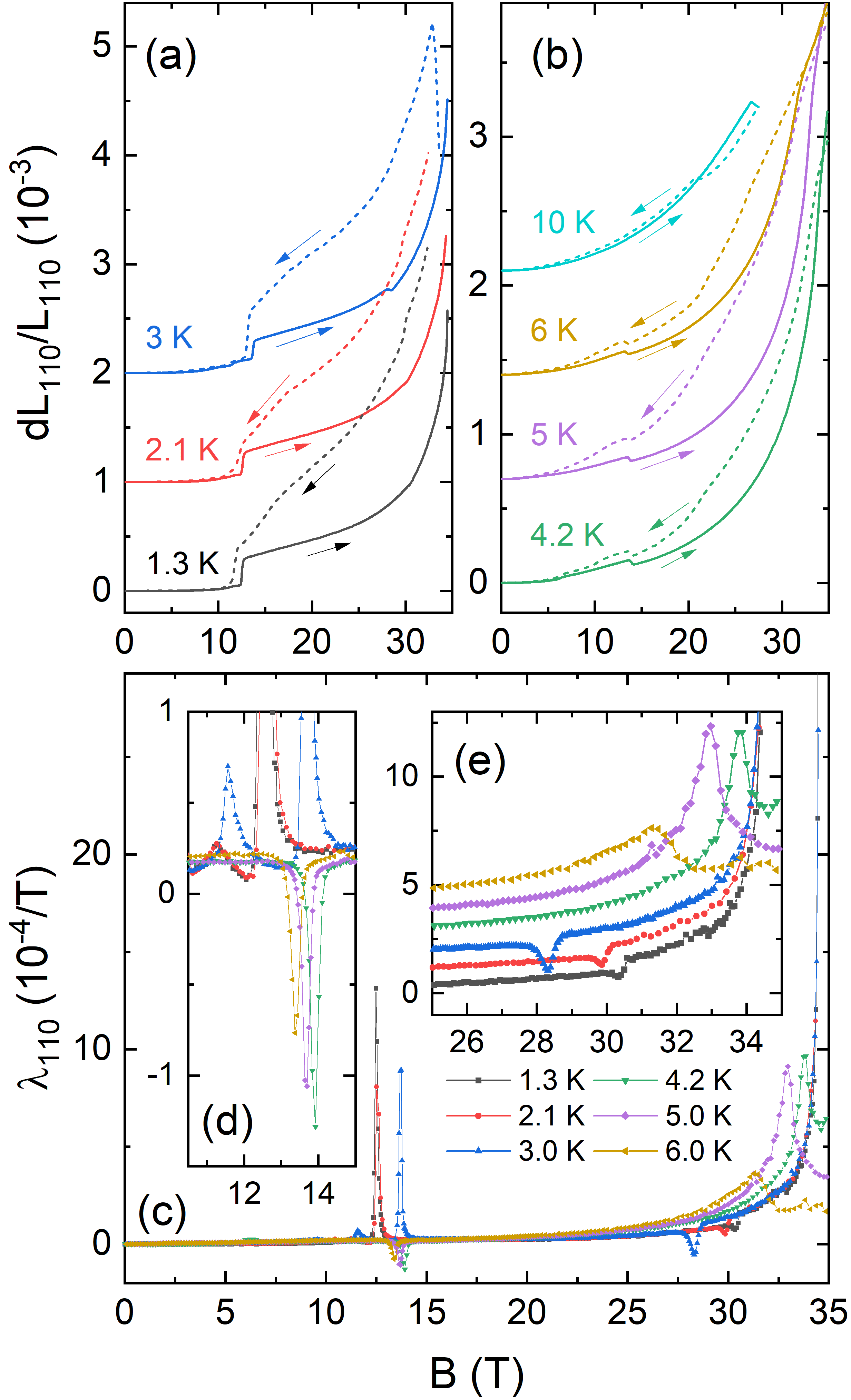}
	\caption{Magnetostriction measurements for $B\parallel [110]$ up to 35~T at selected temperatures. (a-b) Magnetostrictive relative length changes, (c-e) magnetostriction coefficient in different field ranges (up-sweeps only). Data in (a) are offset by 10$^{-3}$ and by $7\times 10^{-4}$ in (b). Dashed lines show down-sweeps as indicated by the arrows. Data in (e) are offset by $8\times 10^{-5}$/T for clarity.} \label{MS_DLL_HFML}
\end{figure}

\begin{figure}[htb]
	\centering
	\includegraphics [width=\columnwidth,clip] {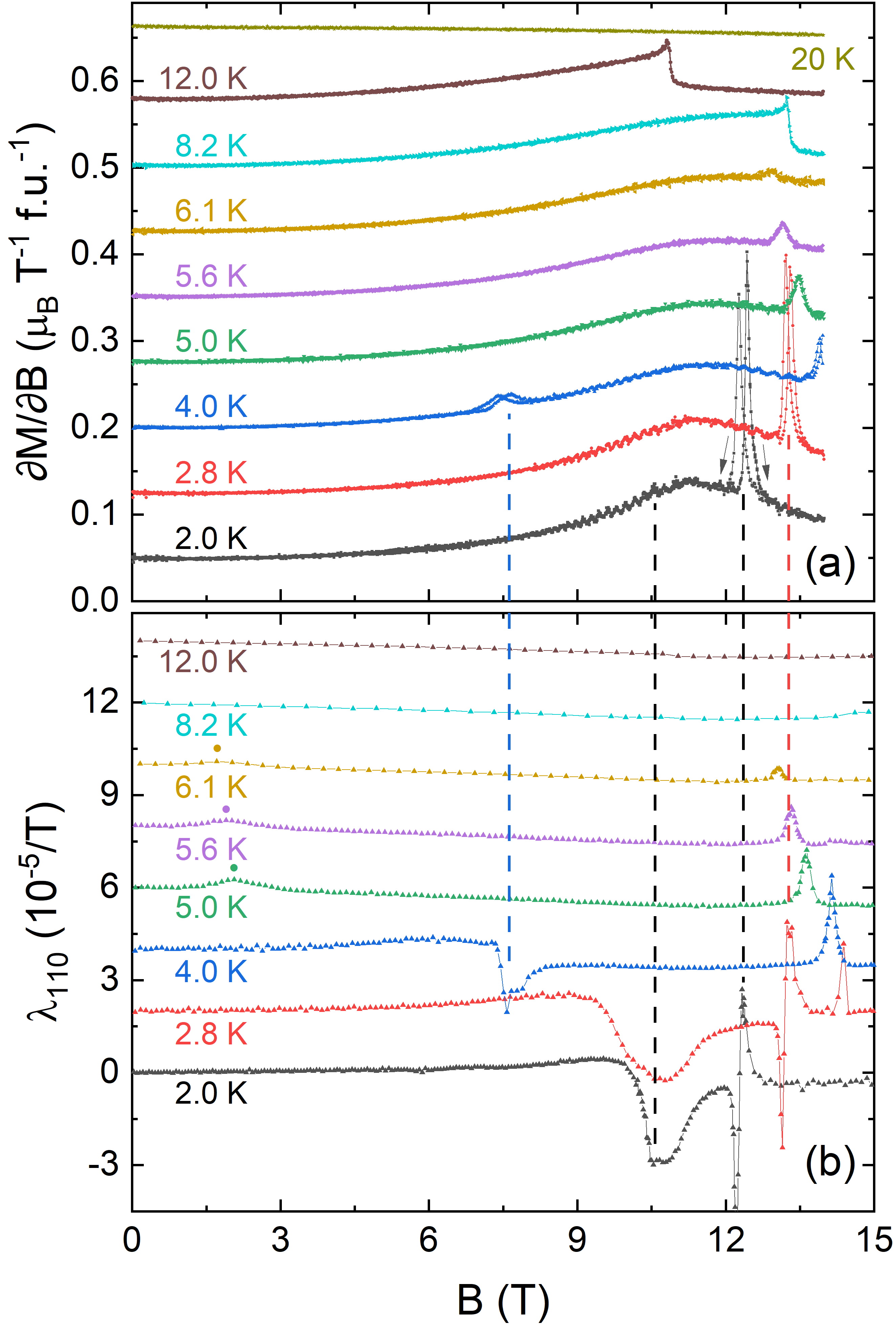}
	\caption{(a) Magnetic susceptibility ${\partial}M/{\partial}B$ and (b) magnetostriction coefficient $\lambda_{\rm [110]}$ for $B\parallel [110]$ at selected temperatures below \TN. Dashed vertical lines emphasize the co-occurrence of anomalies in both quantities. Colored triangles at 5.0~K to 6.1~K data indicate an additional anomaly which is visible in $\lambda_{\rm [110]}$ but not in ${\partial}M/{\partial}B$. Data in (a) are offset by $7.5\times 10^{-2}\mu_{\mathrm{B}}$/(T f.u.) and by $2\times 10^{-5}$/T in (b).} \label{MS_dMdB-vs-Lambda}
\end{figure}

Magnetostriction measurements along the [110] direction with $B\parallel [110]$ show that
the zero field LT and IT phases do not persist up to highest applied magnetic fields but, similar to the case where the magnetic field was applied along the c axis, new phases appear in field (Figs.~\ref{plateau},~\ref{MS_DLL_HFML} and \ref{MS_dMdB-vs-Lambda}).
At low temperatures, two hysteretic discontinuous phase transitions are revealed by jumps in $L(B)$ at $B_{\rm a1} = 11.3$~T (1.3~K) to 11.6~T (3~K) and at $B_{\rm a2} =12.5$~T (1.3~K) to 13.7~T (3~K), as well as a continuous transition around 30~T (1.3~K) to 28.3~T (3~K) (Fig.~\ref{MS_DLL_HFML}(a); see also Fig.~\ref{PD_Both}(c) below). While $B_{\rm a1}$ is barely visible in $M(B)$, $B_{\rm a2}$ is associated with a distinct magnetization jump $\Delta M$ (see Fig.~\ref{plateau}).

In the intermediate temperature phase (which, for $B = 0$~T, appears at 4.8~K $< T < 6.8$~K), several transitions into different phases are observed too as a magnetic field $B\parallel [110]$ is applied.
Firstly, at low fields around 2~T the magnetostriction data show a hysteretic anomaly which is not visible in the magnetization (Fig.~\ref{MS_dMdB-vs-Lambda}).
Increasing the field above 13~T reveals a first order phase transition marked by jumps in $L(B)$ and $M(B)$ (Fig.~\ref{MS_DLL_HFML}(b,d) and \ref{MS_dMdB-vs-Lambda}).
Above 30~T an anomaly is visible which changes shape from a seemingly discontinuous nature at 4.2~K to a more continuous behavior at 6.0~K (Fig.~\ref{MS_DLL_HFML}(e)).
All measurements up to 6~K display a strongly hysteretic behavior in the magnetostriction as the field is decreased from the highest fields to about 10~T (Fig.~\ref{MS_DLL_HFML}(a) and (b)).

\begin{figure}[htb]
	\centering
	\includegraphics [width=\columnwidth,clip] {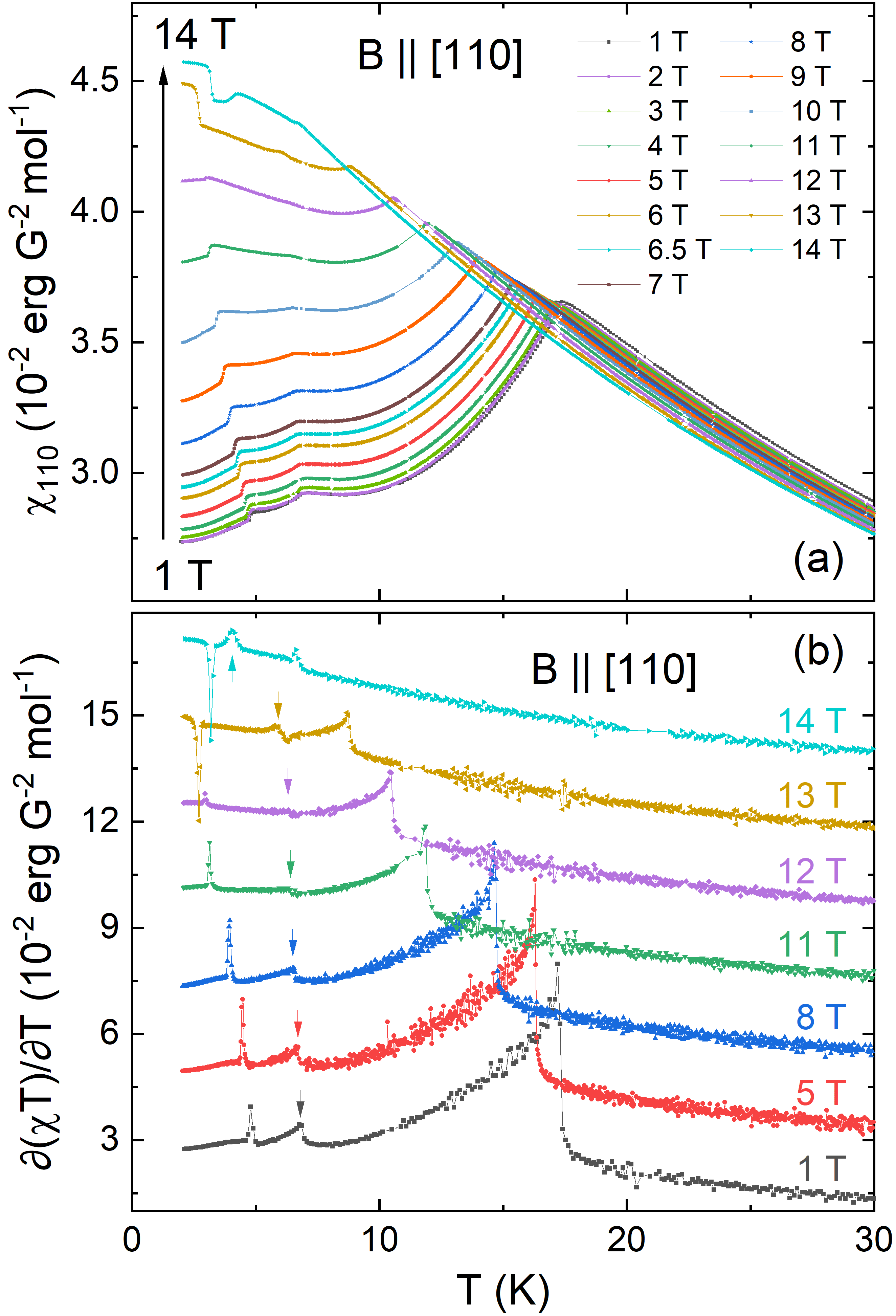}
	\caption{(a) Static magnetic susceptibility $\chi=M/B$ and (b) Fisher's specific heat ${\partial}(\chi T)/{\partial}T$ in magnetic fields $B\parallel [110]$ between 1 and 14~T. Colored arrows indicate $T_{\mathrm{IT}}$ at the respective applied field. The data in (b) are offset by $2.1\times 10^{-2}$ erg/(G$^2$ mol).} \label{110_M-of-T}
\end{figure}

The results from thermal expansion measurements for $B\parallel [110]$ up to 15~T are shown and described in detail in the Supplement Material (Fig.~S1 to S5)~\cite{SM}). Together with measurements of the static magnetic susceptibility in fields up to 14~T (Fig.~\ref{110_M-of-T}(a)) and their derivative in the form of Fisher's specific heat~\cite{Fisher1962}, ${\partial}(\chi T)/{\partial{T}}$, they fully confirm the findings from the magnetostriction measurements.

\begin{figure}[htb]
	\centering
	\includegraphics [width=\columnwidth,clip] {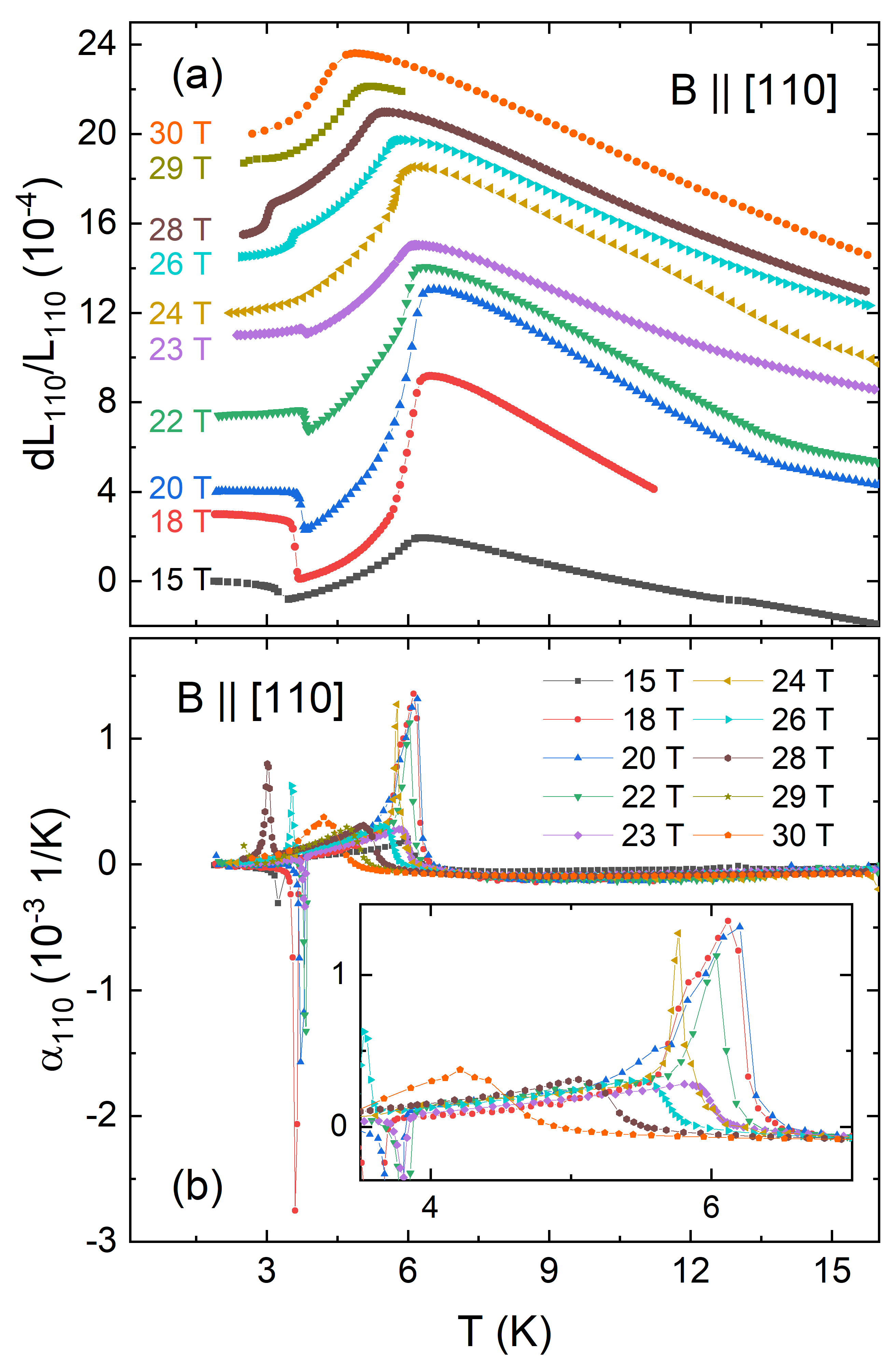}
	\caption{(a) Relative length changes and (b) thermal expansion coefficient at high magnetic fields $B\parallel [110]$ up to 30~T at temperatures between 2 and 16~K. The insert in (b) shows a magnification of the anomalies between 4 and 6~K.} \label{TE_HFML}
\end{figure}

Thermal expansion measurements between 15~T and 30~T reveal two anomalies which can be traced to the highest measured fields (Fig.~\ref{TE_HFML}). At lower temperatures a jump in $L(T)$ clearly signals a discontinuous phase transition. This transition shifts to higher temperatures and the corresponding anomaly shrinks in magnitude as the field is increased to 24~T. Above 24~T it changes sign, shifts back to lower temperatures and grows in magnitude. At 30~T this transition is not visible anymore, at least down to 2.7~K.
The second anomaly around 6~K (at 15~T) is visible as a kink in $L(T)$, signaling its continuous nature. As the magnetic field is increased it continuously shifts to lower temperatures. Noticeably, the measurements at 18, 20, 22 and 24~T show a more jump-like behavior for this transition. These differences and their dependence on the sample history are discussed in the Supplemental Material~\cite{SM}.

\subsection{Phase diagrams}

\begin{figure*}[htb]
	\centering
	\includegraphics [width=2\columnwidth,clip] {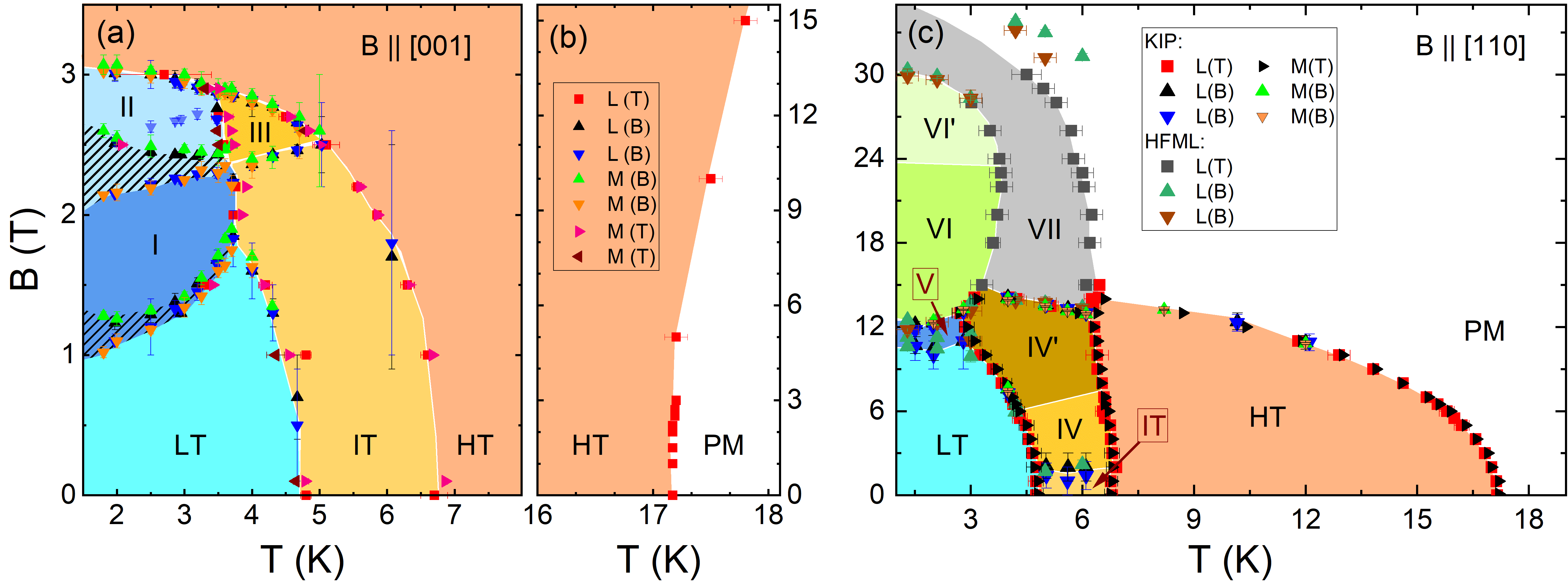}
	\caption{Magnetic phase diagram of \NdB\ for (a) $B\parallel [001]$ at low temperatures up to 3.5~T, (b) $B\parallel [001]$ at the HT-PM phase transition for fields up to 15~T, and (c) for $B\parallel [110]$ up to 35~T. Phase transitions derived from dilatometric measurements using temperature sweeps ($L(T)$) and field sweeps ($L(B)$ up and down sweeps) as well as magnetization measurements ($M(B)$ up/down sweeps; $M(T)$ taken upon warming/cooling) are indicated by the different markers (see legends). Differently colored areas indicate the different phases as described in the main text.} \label{PD_Both}
\end{figure*}

{\boldmath $B\parallel c$:} The phase diagram for magnetic fields up to 15~T applied along the $c$ direction is shown in Fig.~\ref{PD_Both}(a) and (b). It has been constructed from the thermal expansion, magnetostriction and magnetization data presented above. A total of seven distinct phases are found, i.e., one new phase (III) below 3~T was uncovered and phase II is confirmed with respect to the previously reported phase diagrams~\cite{Watanuki2009, Brunt2018}.

The phase boundary which is the highest in temperature, i.e., \tn ($B$), separating the paramagnetic (PM) phase and the HT phase, shifts to higher temperatures for higher fields $B\parallel [001]$ in the field range under study. $\lambda$-like anomalies at \tn ($B$) show the continuous nature of this transition. In contrast to the HT phase, the intermediate temperature phase is suppressed in magnetic field. The change of the shape of the corresponding anomaly at 2.7~T (Fig.~\ref{TE_c}) indicates a change in the phase transition from a continuous to a discontinuous type. This change corresponds to the presence of an additional phase which we label phase III (dark yellow in Fig.~\ref{PD_Both}(a)). Similar to the IT phase, the LT phase is also suppressed as a magnetic field is applied. While the LT-IT phase boundary shows a positive uniaxial pressure dependence marked by positive jumps in the thermal expansion coefficient, the IT-I transition exhibits a negative dependence on $p_c$. The same holds for the phase boundary between phases II and III. All three aforementioned phase transitions (LT-IT, IT-I, II-III) are of discontinuous nature, and the LT-I, I-II, and II-HT transitions as well. Hysteresis at the LT-I and I-II phase boundaries is shown as shaded areas in the phase diagram. Both phase I and phase II correspond to the magnetic field and temperature regime, where plateaus are observed in the magnetization data. While phase I is characterized by the 1/5 magnetization plateau, the 1/4 plateau occurs in the regime of phase II (cf. Fig.~\ref{plateau}). In contrast to the 1/5 plateau, which has been observed previously~\cite{Brunt2018}, the 1/4 plateau in phase II is newly observed here. Downwards pointing triangles in phase II correspond to the additional jumps observed in the magnetostriction data, which are assumed to rather belong to the HT-II phase transition than indicating an additional phase boundary. For fields above 3~T the HT phase prevails down to the lowest measured temperatures.\\

{\boldmath $B\parallel [110]$:} Fig.~\ref{PD_Both}(c) shows the magnetic phase diagram up to 35~T for fields applied along the [110] direction. It was constructed from thermal expansion, magnetostriction and magnetization data. A total of ten phases has been found, unveiling six previously unknown phases~\cite{Watanuki2009, Brunt2018}.


Starting at \TN~$= 17.2$~K in zero-field, the evolution of magnetic order in the HT phase is strongly suppressed in fields up to 14~T. The HT-PM phase boundary is of continuous nature and exhibits a positive uniaxial pressure dependence on $p_{\rm 110}$ (see the Supplemental Material, Figs.~S2 to S5~\cite{SM}). The HT-IT phase boundary as well as the adjacent HT-IV and HT-IV' boundaries~\footnote{We label an arbitrary phase boundary as $T_i(B)$ respectively $B_i(T)$ with $T_i$/$B_i$ the transition temperatures/fields separating the respective phases.} are only weakly field-dependent and mark continuous phase transitions. In contrast to the HT-IV transition temperature which exhibits a negative uniaxial pressure ($p_{110}$) dependence, ${\partial}T_i/{\partial}p_{110}$ is positive at both the HT-IT and HT-IV' transitions. As will be further discussed below, we hence consider IV and IV' distinct thermodynamic phases. All transition temperatures to the LT phase (i.e., LT-IT, LT-IV and LT-IV') are suppressed in the field. The LT-IT phase transition is of discontinuous nature, with ${\partial}$\tit $/{\partial}p_{\rm 110}>0$. A positive pressure dependence is also observed for the LT-IV phase boundary, which is of a continuous type (see the Supplemental Material, Figs.~S2 and S3~\cite{SM}).
The boundary between phases IV' and LT is of discontinuous nature and exhibits negative uniaxial pressure dependence, ${\partial}T_{\rm LT-IV} /{\partial}p_{\rm 110}<0$. The presence of a distinct phase IV and the resulting presence of a phase boundary between IV and IV' is evidenced by the changes in sign and behavior of the corresponding anomalies in $L(T)$ as the magnetic field is increased. Notably, no anomalies are observed in the magnetostriction data at the expected positions. Recalling the Maxwell relation $1/L_i \times (\partial L_i/\partial B)_T = -(\partial M/\partial p_i)_B$ this observation indicates that the uniaxial pressure dependence of the magnetization, ${\partial}M/{\partial}p_{\rm 110}$, does not significantly change at the phase boundary.
A summary of the different phase boundaries and their behavior under uniaxial pressure $p \parallel [110]$ is given in Table~\ref{TE_110_PTs}.

\renewcommand{\arraystretch}{1.2}
\begin{table}[!ht]
\setlength{\tabcolsep}{4pt}
\caption{Summary of phase boundaries determined from dilatometry studies along [110] with $B\parallel [110]$, including their uniaxial pressure dependencies and nature (discontinuous/continuos (d/c)).}
\label{TE_110_PTs}
    \centering
    \begin{tabular}{l| c c}
    \hline \hline
        Phase boundary & ${\partial}T_{\mathrm{crit}}/{\partial}p_{110}$ & Type \\ \hline
        LT-IT & $> 0$ & d \\
        LT-IV & $> 0$ & c \\
        LT-IV' & $< 0$ & d \\
        HT-IT & $> 0$ & c \\
        HT-IV & $< 0$ & c \\
        HT-IV' & $> 0$ & c \\
        HT-PM & $> 0$ & c \\
        VI-VII & $< 0$ & d \\
        VI-IV' & $> 0$ & d \\
        VII-PM & $> 0$ & c \\ \hline \hline
    \end{tabular}
\end{table}
\renewcommand{\arraystretch}{1}

Concerning the transitions driven by magnetic field at the lowest temperatures, it is not completely clear from the data whether phase V can be considered a proper thermodynamic phase or only a hysteresis area between the LT phase and phase VI.
For fields $B||[110] > 12$~T, while the phase boundaries are unambiguously derived from the experiments, characteristics of the anomalies -- whether they are continuous or discontinuous or what kind of pressure dependence they exhibit as well as their magnitude -- vary strongly between measurements with different setups, i.e., differing small but finite pressure applied on the sample. We attribute this to an interplay of domain effects, hysteresis, and the presence of small but finite strain inevitably applied in capacitive dilatometry which is discussed in detail in the Supplement Material~\cite{SM}. Hence, thermodynamic relations cannot be applied to obtain, e.g., uniaxial pressure effects for these anomalies. The positions of the anomalies, however, are consistent throughout all measurements, which allows us to reliably construct the phase diagram presented here even for highest fields. Note, that the sign changes of the anomalies in $L(T)$ separating the phases VI/VI' from VII and the phases IV/IV' from HT have been obtained using a single setup and hence are not affected by domain effects. The fact that the sign change in $\partial T_{\rm VI/VI'-VII}/\partial p_{\rm 110}$ corresponds with a sign change in the magnetization anomaly as evidenced by the slope of $T_{\rm VI/VI'-VII}(B)$ strongly corroborates the conclusion of two distinct phases VI and VI'.

\section{Discussion}


The effect of uniaxial pressure on discontinuous phase transitions is associated with the jumps in relative length ($\Delta L/L$) and in entropy ($\Delta S$) at the phase boundary via the Clausius-Clapeyron-relation which in the case of $T_{\rm LT}(B)$ reads:

\begin{equation}
\frac{\partial T_{\rm LT}}{\partial p_i}=V_m\frac{\Delta L_{i,{\rm LT}}/L_i}{\Delta S_{\rm LT}}.
\end{equation}

Due to general experimental uncertainties when measuring specific heat anomalies at first order phase transitions by means of the relaxation method, we have determined the entropy changes at \tlt ($B||c$) by exploiting the slope of the phase boundary $\partial T_{LT}/\partial B$ and the measured jumps in the magnetization: $\Delta S_{\rm LT}= (\partial T_{\rm LT}/\partial B)^{-1}\times\Delta M_{\rm LT}$~\cite{Spachmann2021}. The resulting entropy jumps and pressure dependencies at 1, 2 and 2.5~T as determined from our data are summarized in Table~\ref{pressure}. Specifically, these values characterize the pressure dependencies of the IT-LT (1~T), IT-I (2~T) and II-III (2.5~T) phase boundaries, respectively. The latter two are negative, whereas the IT-LT transition shows a positive dependence on pressure applied along the $c$ axis. We also note the much larger values for the IT-I transition which is one order of magnitude larger than the other two, i.e., phase I is strongly suppressed by the uniaxial pressure $p \parallel c$.


\renewcommand{\arraystretch}{1.5}
\begin{table}[htb]
\setlength{\tabcolsep}{3pt}
\caption{Relative length changes, jumps in magnetization, and field dependence of the critical temperature at different magnetic fields $B\parallel c$ as well as (1) derived jumps in entropy and (2) pressure dependence of the ordering temperature for uniaxial pressure along the crystallographic $c$ axis. The associated phase boundaries are given in column 1.}
\label{pressure}
\begin{center}
\begin{tabular}{c |c | c c c c c }
		
\hline\hline

phase& $B$ & $\frac{\Delta L_c}{L_c}$ & $\Delta M$ & ${\partial}T/{\partial}B$ & $\Delta S$&${\partial}T/{\partial}p_{c}$\\
bound.&(T)&$(10^{-6})$&($10^{-3}\frac{{\rm \mu_B}}{\rm f.u.}$)                & (K/T)    & (J/K)           &($\frac{\rm K}{\rm GPa}$)\\
\hline
IT-LT & 1   & 2.3(3)& 14.7(3) & 0.30(8) & 0.3(1) & 0.25(4)\\
IT-I&2 & -7(1)&24(2)           & 0.85(8)     & 0.16(3) & -1.4(5)\\			
II-III&2.5 & -7.2(7)& 44(2)        &0.20(8) & 1.2(5)       & -0.19(10)\\
\hline\hline
		
\end{tabular}	
\end{center}
\end{table}
\renewcommand{\arraystretch}{1}

The phase boundary \tn ($B$) between PM and HT phases depends only very weakly on $B \parallel c$ whereas it is strongly suppressed by $B \parallel [110]$ (see Fig.~\ref{PD_Both}). When associated with a decrease of the magnetization as observed for $B \parallel [110]$, long-range AFM order is indeed expected to be suppressed in magnetic fields. Hence, the effect of $B \parallel [110]$ resembles the typical behavior of an antiferromagnetic phase. This is supported by neutron diffraction data, which suggest an all-in/all-out arrangement of the spins for this phase~\cite{Metoki2018,Yamauchi2017}.

\begin{figure}[htb]
	\centering
	\includegraphics [width=\columnwidth,clip] {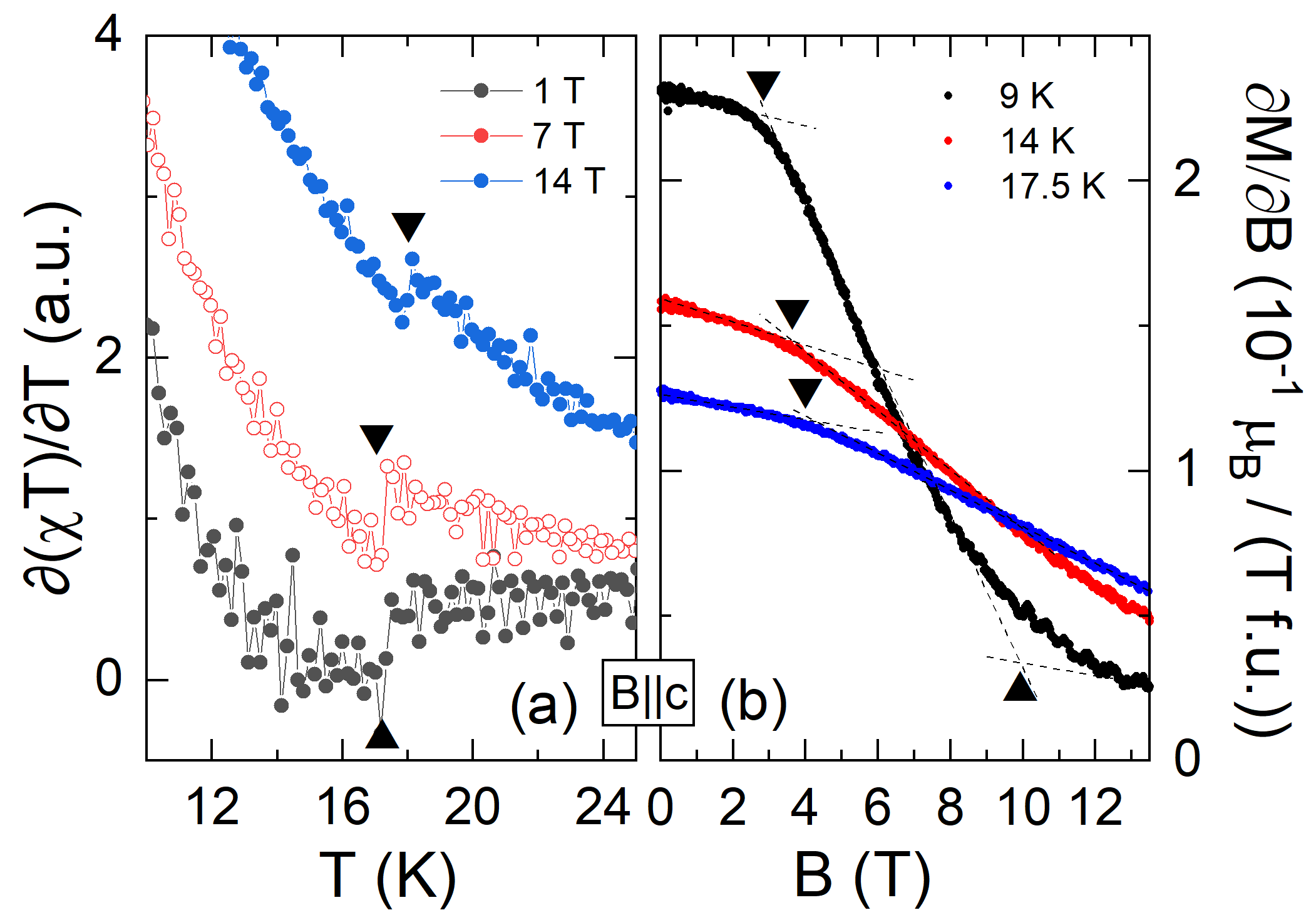}
	\caption{(a) Fisher's specific heat at 1, 7, and 14~T with $B\parallel c$. Triangles indicate \TN ($B$). (b) Magnetic susceptbility $\partial M/\partial B$ at 9, 14, 17.5 K. Triangles indicate a change of behavior in $M$ vs. $B$ (see the text).} \label{M_c}
\end{figure}

Correspondingly, the constant behavior of \tn\ at low fields $B \parallel c$ agrees well to the fact that no anomalies had previously been observed in magnetic susceptibility measurements~\cite{Brunt2019,Brunt2018,Watanuki2009} in this field range, which may be explained by the easy magnetic axis being $\perp c$ or by easy plane anisotropy in the cAFM HT phase~\cite{Metoki2018,Yamauchi2017}. However, a small positive jump in the derivative of our susceptibility measurements signals the magnetic transition already at small field and this feature becomes more pronounced for higher fields (Fig.~\ref{M_c}(a)). Notably, the anomaly signals larger magnetization in the ordered phase as compared to the PM phase which contradicts the expectations for an ordinary AFM phase. In addition, \tn ($B$) shifts to higher temperatures for $B \parallel c\gtrsim 6$~T. This again clearly contradicts a simple AFM behaviour and agrees to the sign of the anomaly in $M(T)$ as well as to the observation that the kink in the magnetization at \tn ($B$) becomes more pronounced at 7 and 14~T (Fig.~\ref{M_c}(b)). Hence, both the anomaly sign and the slope $\partial T_{\rm N}/\partial B_{|| c}>0$ of the phase boundary clearly imply the evolution of an increasing-in-field ferromagnetic component in the HT spin structure for $B \parallel c\gtrsim 6$~T.

\begin{figure}[htb]
	\centering
	\includegraphics [width=\columnwidth,clip] {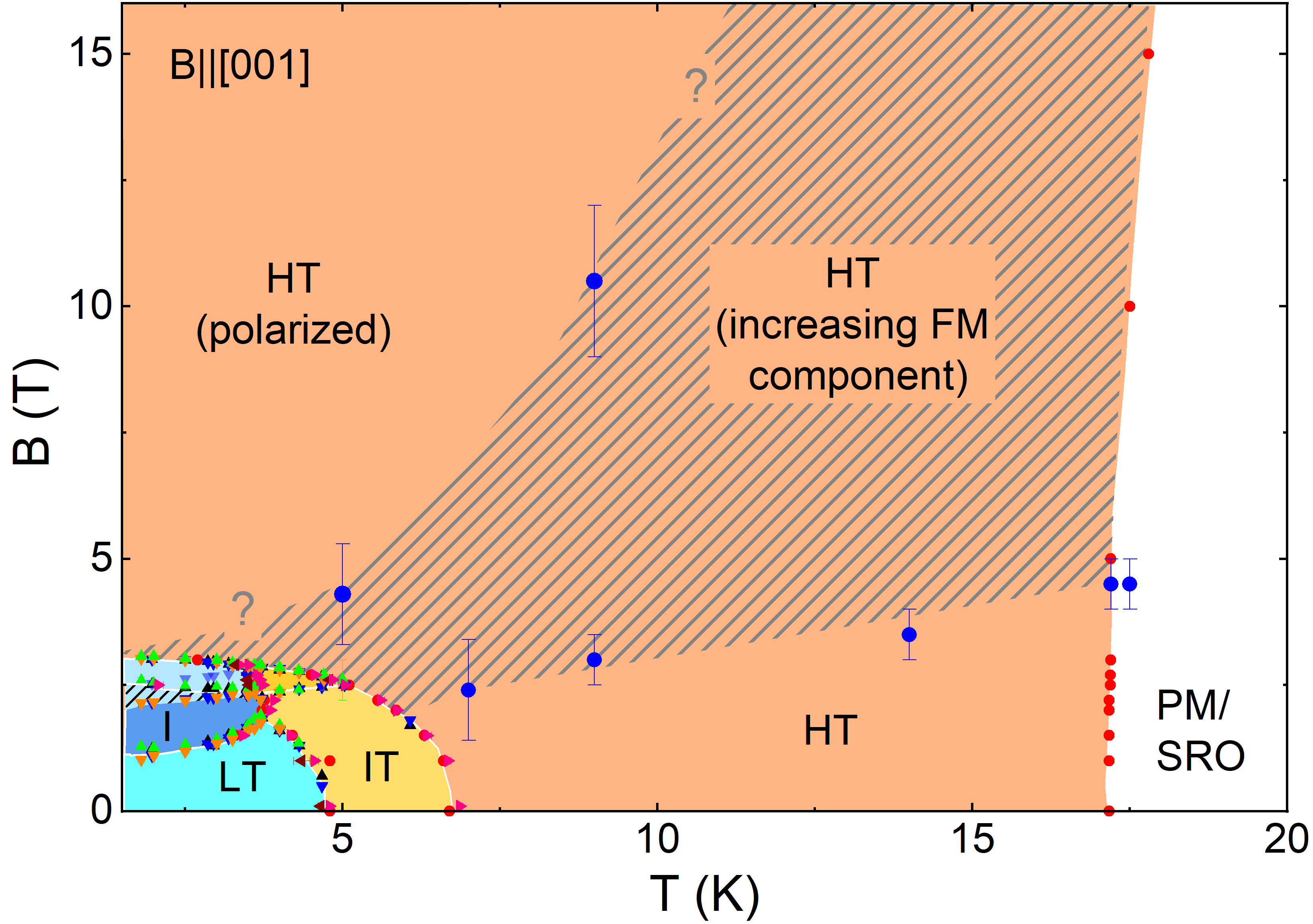}
	\caption{Magnetic phase diagram of \NdB\ for $B\parallel c$ highlighting the behaviour in the HT phase. The bottom left part reproduces Fig.~\ref{PD_Both}(a). The grey shaded area is a guide to the eye marking the regime of field-driven increase of a small ferromagnetic moment (see the text); blue markers confining this region correspond to the triangles in Fig.~\ref{M_c}.} \label{phd-c}
\end{figure}

The data hence show gradual polarization of magnetic moments along the magnetic field direction as may be expected in case of easy-plane anisotropy. Indeed, $M(B)$ curves in the HT phase (Fig.~\ref{M_c}(b)) imply a quasi-linear behaviour at small fields while the slope changes towards gradual right-bending indicative of an increasing ferromagnetic moment before tending to saturation. This behaviour is similar as, e.g., observed for weak ferromagnetic moments in La$_{5/3}$Sr$_{1/3}$NiO$_4$~\cite{Klingeler2005}. The regime of increasing ferromagnetic moment is marked by the shaded area in Fig.~\ref{phd-c} which neighbors the region of the phase diagram with $\partial T_{\rm N}/\partial B>0$. Whether the increase of \tn\ in magnetic fields is directly associated with the presence of a small spin component of the afm ordered moment along the $c$ axis of about $m_c \simeq 0.2$~\mb~\cite{Metoki2018} cannot be concluded from our thermodynamic data. We however note the absence of a potential jump in magnetization which would reflect a hypothetical spin-flip of the small moments, as, e.g., observed in BaCu$_2$Si$_2$O$_7$~\cite{Tsukada2001}.
The kink in $\partial M/\partial B$ which is, e.g., observed at 9~K and roughly 3~T (Fig.~\ref{M_c}(a)), corroborates the notion of a gradual change in the spin order rather than a distinct phase boundary.

A further distinct feature of the phase diagram for $B||{\rm [110]}$ in Fig.~\ref{PD_Both}(c) is the change in slope between the ordered phases and the PM phase at 6~K and 14~T. In this region, \tn ($B$) is very flat and exhibits $\partial T_{\rm N}/\partial B \simeq -2.6(5)$~K/T. Extrapolating the critical field to low temperatures would yield the critical field $B_{\rm c}(T=0~{\rm K})\simeq 15$~T. In contrast, the phase boundary between the PM/HT phase and the sequence of field induced phases IT-IV-IV'-VII is steep as $\partial B_i/\partial T\simeq -30(5)$~T/K at $B\simeq 14$~T. We note however that the phase diagram for $B>14$~T has been constructed from dilatometric studies so that potential phases which are only very weakly affected by uniaxial pressure might be missed (see Fig.~S13 in the SM~\cite{SM}). For $B\leq 14$~T, our  magnetization and specific heat data exclude further phase boundaries. However, independent on this possibility, the above mentioned difference in the slopes can be quantitatively evaluated using the Ehrenfest equation:
\begin{equation}
    \frac{\partial T_{\rm i}}{\partial B}=-T_{\rm i}\frac{\Delta M'}{\Delta c_{\rm p}}.\label{ehrenfest}
\end{equation}
Here, $T_{\rm i}$ is the transition temperature, $\Delta c_{\rm p}$ the associated jump in specific heat, and $\Delta M'=\Delta (\partial M/\partial T)$ the associated change in slope of magnetization. Employing a method of equal-area construction to determine $\Delta M'$ from the data~\cite{Gries2022}, 
Eq.~\ref{ehrenfest} allows us to estimate the specific heat changes $\Delta c_{\rm p}$ at the phase transitions. At \tn ($B=13$~T), we obtain $\Delta c_{\rm p}\simeq 0.4(1)$~\jmk . This small value agrees to the assumption that in high magnetic fields the onset of AF order does not consume significant amounts of entropy. In contrast, we find $\Delta c_{\rm p}\simeq 9(2)$~\jmk\ at the PM-VII phase boundary. Notably, this value is even slightly larger than $\Delta c_{\rm p}(B=0~{\rm T})\simeq 5(1)$~\jmk\ determined by specific heat measurements at \tn\ in zero magnetic field~\cite{Ohlendorf2021}. This result implies that entropy changes at $T_{\rm PM-VII}$ are in stark contrast to the ones at \tn ($B\simeq 13$~T) and we conclude that additional degrees of freedom beyond spin are relevant. A candidate are orbital-related entropy changes as orbital degrees of freedom are relevant in NdB$_4$~\cite{Watanuki2009,Yamauchi2017,Ohlendorf2021}.

\section{Conclusions}
High-capacitance dilatometry studies supported by specific heat and magnetization measurements were used to construct the magnetic phase diagrams of \NdB\ for $B \parallel c$ and $B \parallel$~[110]. In total, seven new phases have been discovered and we find a 1/4 magnetization plateau in addition to the known 1/5 one. Uniaxial pressure dependencies along the $c$ axis were determined for the IT-LT, IT-I and III-II phase boundaries. For magnetic fields applied along the $c$ axis, the increase of the critical temperature with increasing applied magnetic field for the phase transition between the PM phase and the HT AFM phase suggests that the magnetic moments of the all-in/all-out spin structure of the AFM phase are driven towards the $c$ axis with increasing magnetic field. The results show that capacitance dilatometry is a state-of-the-art method to map out phase diagrams not only due to its high resolution but also because of its sensitivity towards type and sign of the anomalies, which provides further information on phase changes.

\section{Acknowledgments}
We acknowledge financial support by Deutsche Forschungsgemeinschaft (DFG) under Germany’s Excellence Strategy EXC2181/1-390900948 (the Heidelberg STRUCTURES Excellence Cluster) and through project KL 1824/13-1. We also acknowledge support by the European Union’s Horizon 2020 Research and Innovation Programme, under Grant Agreement no. 824109 (European Microkelvin Platform), and by HFML-RU/NWO-I, member of the European Magnetic Field Laboratory (EMFL). Work at the University of Warwick was supported by EPSRC, UK, EP/T005963/1.

\bibliography{2_NdB4_Bibliography}

\end{document}